\documentclass[a4paper]{article}

\usepackage{fullpage,url,amsmath,amsfonts,amssymb,tedmath}

\newcommand{\CC}{\mathcal{C}}

\title{Convolutional codes from units in  matrix and group rings}
\author{Ted Hurley}
\date{}
\begin{document}
\maketitle

\begin{abstract}

A general method for constructing  
convolutional codes from units in Laurent series over matrix rings  is 
presented. Using group rings
as matrix rings, this forms a basis for in-depth exploration of
convolutional codes from group
ring encodings, wherein the ring in the group ring is itself a group
ring. The method is used to algebraically construct series of convolutional
codes. Algebraic methods are used to compute free distances and to
construct convolutional codes to prescribed distances.

\end{abstract}

\section{Introduction}

Methods are presented for constructing convolutional codes using units 
in  Laurent series of finite support over
matrix rings. By  considering group rings as
matrix rings, convolutional codes are constructed from 
units in Laurent series over group rings; these may be considered as group
rings over group rings. Thus convolutional codes are constructed 
 by considering a group ring $RG$ where the ring $R$ is itself a group ring. 

The methods are based on the general method in \cite{hur1} for
constructing {\em unit-derived} codes from group rings where now the
ring of the group ring is a group ring and the group of the
group ring may be  an infinite group such as the infinite cyclic group.

For general information on group rings and related algebra see \cite{seh}.

Using these algebraic methods, the range of convolutional codes
available is expanded and series of convolutional codes are
derived. Free distances and  codes to a prescribed free distances 
may also be derived. Indeed many of the existing convolutional codes can be
obtained in the manner of this paper. 

The paper \cite{mac} is an often quoted source
of information on convolutional codes wherein is  mentioned the lack of
algebraic methods for constructing convolutional codes; and that many of the
existing ones have been found by computer search and are of necessity
of relatively short memories.  

The methods are fairly general  and use properties of group rings
and their embedding into matrix rings. 
Zero-divisors and units in group rings enables the construction
of units in certain polynomial rings and/or group rings over these group rings
from which convolutional codes can be constructed.  Properties of the
convolutional codes can
 be studied and derived from properties of group rings. In many instances the
free  distances can be calculated
algebraically and convolutional codes to a specified free distance, as
for example in \thmref{best} or \thmref{hamming} below, can
be constructed.

The following are some  of the applications of the general
method and these in themselves constitute new methods for constructing
convolutional codes:
\begin{itemize} \item The construction of  series of binary 
$(2,1)$ convolutional
  codes and calculation of their free distances using the group ring
  $(FC_2)C_\infty$ where $F$ is a field of characteristic $2$;
\item Given a linear cyclic code $\CC$  
  with $d = \min(d_1,d_2)$ where $d_1$ is the minimum distance of $\CC$
  and $d_2$ is the distance of the dual of $\CC$, the 
  generator polynomial $f$ of $\CC$ is mimicked in $RC_\infty$ 
to construct  convolutional
  $(2,1)$ codes of minimum free distance $d+2$; \item The construction
  of  
rate $\frac{3}{4}$ and higher rate convolutional codes with
  prescribed minimum distance; \item The construction of convolutional codes
  over a field $F$ of characteristic $p$ for any prime $p$ using nilpotent
  elements in the field $FG$ where $G$ is a group whose order is
  divisible by $p$; \item The construction of  {\em Hamming type}
  convolutional codes and calculating their free distances;
  the  construction of  Hamming-type convolutional codes to a desired minimum
  free distance; \item The construction of convolutional codes using 
idempotents in group rings. These are particularly used in cases where
  the characteristic of the field does not divide the order of the
  group;  { \em characters} of groups and {\em character tables} come
  into play in constructing these convolutional codes.
\end{itemize}

\subsection{Algebraic Description of Convolutional Codes}

Background on general algebra and group rings
may be obtained in \cite{seh}.

For any ring $R$, $R[z]$ denotes the polynomial ring with coefficients
from $R$ and 
$R_{r\times n}$ denotes the ring of $r\times n$ matrices with
coefficients from  $R$.
$R^n$ is used to denote $R_{1\ti n}$ and thus $R^n = \{(r_1, r_2, \ldots,
r_n) : r_i \in R\}$.

It is easy to verify that  $R_{r\times n}[z] \cong
R[z]_{r\times n}$. 

 $R[z,z^{-1}]$ is used to denote the set of Laurent series of finite
support in $z$  with coefficients from $R$. 
{\em Finite support} means that only a finite
number of the coefficients are non-zero. It is clear that $R[z,z^{-1}]
\cong RC_\infty$, where $C_\infty$ denotes the infinite cyclic group. 
(Elements in group rings have finite support.)

Note also the  relationship between $R[z]$
and $RC_\infty$   -- $R[x] \cong T$ where
$T$ denotes the
algebra of {\em non-negative elements}, i.e.\ the algebra 
of elements $w = \di\sum_{i=0}^\infty\al_ig^i$,  in
$RC_\infty$.

If $\F$ is an integral domain then  $\F[z]$ has no zero-divisors
and only trivial units -- the units of $\F[z]$ 
are the units of $\F$.

See \cite{mac} and/or \cite{blahut} for basic information on convolutional
codes and algebraic descriptions are
described therein. 
The (equivalent) algebraic description given  in \cite{heine} is
extremely useful and is given below. 

A convolutional code $\CC$ of length $n$ and dimension $k$ is a direct
summand of $\F[z]^n$ of rank $k$. Here $\F[z]$ is the polynomial
ring over $\F$ and  $\F[z]^n = \{(v_1, v_2, \ldots,
v_n) : v_i \in \F[z]\}$. 

Suppose $V$ is a submodule of $\F[z]^n$ and
that $\{v_1, \ldots, v_r\} \subset \F[z]^n$ forms a generating set for
$V$. Then  $V =
\text{Image}  \, M = \{uM : u \in \F[z]^r\}$ where 
$M = \left[\begin{array}{l}v_1 \\ \vdots \\ v_r \end{array}\right] \in
\F[z]_{r\times n}$. This $M$ is called a {\em generating matrix} of $V$. 

A generating matrix $G \in \F[z]_{r\times n}$ having rank $r$ is called a
{\em generator} or {\em encoder matrix}  of $\CC$. 

A matrix $H \in \F[z]_{n\times(n-k)}$ satisfying $\CC = \ker H =
\{v \in \F[z]^n : vH = 0 \}$ is said to be a {\em control matrix} of the code
$\CC$.

\section{Convolutional codes from units}\label{sec:units}

Let $R$ be a ring which is a
subring of the  ring of matrices $F_{n\ti n}$. 

 In particular the   
group ring $FG$  is a subring of $F_{n\ti n}$, where $n = |G|$, by
an explicit embedding given in \cite{hur2}. 
There is no restriction
on $F$ in general but it is assumed  to be a field here; however many
of the results will hold more generally. 

{\em Units} and {\em zero-divisors} in any ring are defined in the usual manner. 

Construct $R$-convolutional codes  as follows:

\subsection{Polynomial case}\label{sec:units1} For clarity the polynomial case is
considered initially although this is a special case of the more
general construction. 

Suppose $f(z) g(z) = 1 $ in $R[z]$. Essentially then  the encoder matrix 
  is obtained from  $f(z)$ and the decoder or control matrix is  
  obtained from  $g(z)$ using 
  a variation on the method for constructing unit-derived 
  codes as formulated in \cite{hur1} for  non-singular matrices.

Now $f(z) = (f_{i,j}(z))$ is an $n\times n$ matrix with
  entries $f_{i,j}(z) \in F[z]$. Similarly $g(z) = (g_{i,j}(z))$ is an
  $n\times n$ matrix over $F[z]$. 
Suppose $r[z] \in F[z]^r$ and  consider $r[z]$ as an element of
  $\F[z]^n$ (by adding zeros to the
end of it).  Then define a mapping  $\gamma: 
F[z]^r \rightarrow
  F[z]^n$ by 
$\gamma: r(z) \mapsto r(z) f(z)$.  The code $\CC$ is the image of
  $\gamma$. Since $r[z]$ has
  zeros in its last $(n-r)$ entries as a member of $F[z]^n$, this means
  that {\em the generator matrix is the first $r$ rows} of $f(z)$
  which is an $r\times n$ matrix over $F[z]$. Since $f(z)$ is
  invertible, this generator matrix has $\rank r$ and is thus the
  encoder matrix which we denote  by  $G(z)$. For this polynomial
  case, $G(z)$
  is a basic generator matrix -- see A.1 Theorem in \cite{mac}. 

$w(z) \in \F[z]^n$ is a codeword if and only if $w(z) g(z)$ is in
$\F[z]^r$, that is,  if and only if the final $(n-r)$ entries of
  $w(z)g(z)$   are
all $ 0$. 
 Suppose $w(z) = (\al_1(z), \al_2(z), \ldots, \al_n(z))$. Then this
 condition is that 

$$   (\al_1(z), \al_2(z), \ldots,
   \al_n(z))*\left(\begin{array}{lllll}g_{1,r+1}(z) &
   g_{1,r+2}(z) & \ldots & g_{1,n}(z) \\ g_{2,r+1}(z) & g_{2,r+2}(z) &\ldots &
   g_{2,n}(z) \\ \vdots &\vdots & \vdots & \vdots \\ g_{n,r+1}(z) &
   g_{n,r+2}(z) & \ldots & g_{n,n}(z)\end{array}\right) = 0 $$

The  check or {\em control matrix} $H(z)$ of the code  is thus:

$$\left(\begin{array}{lllll}g_{1,r+1}(z) &
   g_{1,r+2}(z) & \ldots & g_{1,n}(z) \\ g_{2,r+1}(z) & g_{2,r+2}(z) &\ldots &
   g_{2,n}(z) \\ \vdots &\vdots & \vdots & \vdots \\ g_{n,r+1}(z) &
   g_{n,r+2}(z) & \ldots & g_{n,n}(z) \end{array}\right)$$   
 
This has size $n\times (n-r)$ and  is the matrix consisting of the
last $(n-r)$ columns of $g(z)$ or in other words the matrix obtained
by deleting the first $r$ columns of $g(z)$.

Since $f(z), g(z)$ are units, it is automatic that 
 $\rank G(z) = r$ and $ \rank H(z) =
(n-r)$.

\subsubsection{Restatement of polynomial case} Suppose then 
$f(z) g(z) = 1 $ in $R[z]$. 
The set-up may  be restated  as follows:
 
$$f(z) = \left(\begin{array}{c} f_1(z) \\ f_2(z) \end{array}\right)$$

$$g(z) = \left(\begin{array}{rr} g_1(z), & g_2(z)\end{array}\right)$$

where $f_1(z)$ is an $r\ti n$ matrix, $f_2(z)$ is an $(n-r) \ti n $
matrix, $g_1(z)$ is an $n\ti r$
matrix and $g_2(z)$ is an $n\ti (n-r)$ matrix. 

Then $f(z)g(z) = 1$
implies $$ \left(\begin{array}{c} f_1(z) \\ f_2(z) \end{array}\right) \ti 
\left( \begin{array}{cc}g_1(z), & g_2(z)\end{array}\right) = 1$$

Thus $$\left(\begin{array}{cc} f_1(z)g_1(z) & f_1(z)g_2(z) \\
  f_2(z)g_1(z) & f_2(z)g_2(z)\end{array}\right) = 1$$

From this it follows that \\ 
$f_1(z)g_1(z) = I_{r\ti r}, \\ f_1(z)g_2(z) =
0_{r\ti (n-r)}, \\ f_2(z)g_1(z) = 0_{(n-r)\ti r}, \\ 
f_2(z)g_2(z) =
  I_{(n-r)\ti (n-r)}$.

Thus $f_1(z)$ is taken as the generator or encoder  matrix and $g_2(z)$
is then the check or control matrix.
Note that both  $f_1(z),f_2(z)$ have 
 right finite support inverses and thus by  Theorem 6.3 of \cite{mac} the
 generator matrix $f_1$ is noncatastrophic. 

Given $f(z)g(z)= 1$ by the general described  
method of  unit-derived code in 
\cite{hur1} a convolutional code can be constructed 
using {\em any} rows  of $f(z)$. If rows $\{j_1, j_2, \ldots, j_r\}$ are chosen
from $f(z)$ then we get an encoding $F^r[z] \rightarrow F^n[z]$ with generator
matrix consisting of these $r$ rows of $f(z)$ and check/control matrix
is obtained by deleting the  $\{j_1, j_2, \ldots, j_r\}$ columns of $g(z)$.

Cases with  $f(z) g(z) = 1 $,  $f(z), g(z) \in R[z,z^{-1}]$,
will also in a similar manner produce convolutional codes. The next
section, \sref{pow2}, describes the similar process for these  in detail.   

\subsection{More generally}\label{sec:pow2} 

Let $f(z,z^{-1}), g(z,z^{-1}) \in R[z,z^{-1}]$ be such that 
$f(z,z^{-1}) g(z,z^{-1}) = 1$. 

Suppose now $$f(z,z^{-1}) = \left(\begin{array}{c} f_1(z,z^{-1}) \\
  f_2(z,z^{-1}) \end{array}\right)$$
 
$$g(z,z^{-1}) = \left(\begin{array}{rr} g_1(z,z^{-1}), &
  g_2(z,z^{-1})\end{array}\right)$$ 

where $f_1(z,z^{-1})$ is an $r\ti n$ matrix, $f_2(z,z^{-1})$ is an 
$(n-r) \ti n $ matrix,
  $g_1(z,z^{-1})$ is an $n\ti r$
matrix and $g_2(z,z^{-1})$ is an $n\ti (n-r)$ matrix.

Then 
$$ \left(\begin{array}{c} f_1(z,z^{-1}) \\ f_2(z,z^{-1})
\end{array}\right) \ti  
\left( \begin{array}{rr}g_1(z,z^{-1}), & g_2(z,z^{-1})\end{array}\right) = 1$$

Thus $$\left(\begin{array}{cc} f_1g_1 & f_1g_2 
  \\ f_2g_1 & f_2g_2\end{array}\right) = 1$$

From this it follows that \\ $f_1(z,z^{-1})g_1(z,z^{-1}) = I_{r\ti r}$,
\\ $f_1(z,z^{-1})g_2(z,z^{-1}) =
0_{r\ti (n-r)}$,
\\ $f_2(z,z^{-1})g_1(z,z^{-1}) = 0_{(n-r)\ti r}$, 
\\ $f_2(z,z^{-1})g_2(z,z^{-1}) =
  I_{(n-r)\ti (n-r)}$.

Thus $f_1(z,z^{-1})$ is taken as the generator or encoder  matrix and 
$g_2(z,z^{-1})$ is then the check or control matrix.
It is seen in particular  that $f_1(z,z^{-1}),f_2(z,z^{-1})$ have 
 right finite support inverses and thus  by Theorem 6.6 of \cite{mac} the
 generator matrix $f_1$ is noncatastrophic. 

Given $f(z,z^{-1})g(z,z^{-1})= 1$ by the general described method 
of  unit-derived code of
\cite{hur1} codes 
{\em any} rows  of $f(z,z^{-1})$ can be used to construct a
convolutional. 
If rows $\{j_1, j_2, \ldots, j_r\}$ are chosen
from $f(z,z^{-1})$ then an encoding $F^r[z] \rightarrow F^n[z]$ 
is obtained with generator
matrix consisting of these $r$ rows of $f(z)$ and check/control matrix
obtained by deleting the  $\{j_1, j_2, \ldots, j_r\}$ columns of $g(z)$.

\subsubsection{Particular case}

Suppose $f(z) g(z) = z^t $ in $R[z]$. Then $f(z)(g(z)/z^t) = 1$. Now
   $(g(z)/z^t)$ involves negative powers of $z$ but has
  finite support. The encoder matrix 
  is obtained from  $f(z)$ and the decoder or control matrix is  
  obtained from  $(g(z)/z^t)$ using the  
  method  as formulated  in \sref{pow2}. It is also possible to
  consider $(f(z)/z^i)(g(z)/z^j) = 1$ with $i+j=t$ and to derive the
  generator matrix from $(f(z)/z^i)$ and the check/control matrix from
  $(g(z)/z^j)$.

The control matrix contains negative powers of $z$ but a polynomial
control matrix is easy to obtain from this.

Note that $z^{-n}$ are  units  
worth considering in $R[z,z^{-1}]$ but that  other elements in $R[z,z^{-1}]$ 
 may have inverses with infinite support and the inverses are thus outside
 $R[z,z^{-1}]$.  However in some cases $\di\sum_{i=-t}^m \al_i z^i \in
 R[z,z^{-1}]$ has an inverse in $R[z,z^{-1}]$, for example in certain
     cases when the $\al_i$ are nilpotent, and here  also
     convolutional codes may be defined with (direct) noncatastrophic generator
     matrices.  All these are cases of
     $f(z,z^{-1})\ti g(z,z^{-1}] = 1 \in R[z,z^{-1}]$ but may be worth
       considering  originally from polynomials for the construction.

\subsubsection{Uninteresting zero-divisors} In \cite{hur1} units and
zero-divisors in group rings are used to construct codes.
Zero-divisors in $R[z]$ are not too interesting:
Suppose $uw = 0$ in $R[z]$ and $u$ is an element of least degree so
that $uw = 0$. Then $w$ or $u$ has degree zero; if $w$ has degree 0 then it 
 is a zero-divisor of each coefficient of $u$ and if $u$ has degree zero
then it is a zero-divisor of each coefficient of $w$.

Thus if we require zero-divisor codes in $R[z]$ we are looking at 
direct sums of zero-divisor codes in $R$.
Using units  in $R[z]$ to construct codes is far  more productive.

\subsection{Group ring matrices}

In the constructions of \sref{units1} or in the more general
\sref{pow2}, $R$ is a subring
of $F_{n\times n}$.
Suppose now  $R = FG$ is the  group ring of the group $G$ over $F$.
 
 The group ring $RG$ is a subring of  $F_{n\times
  n}$ using an  explicit  correspondence between  the group ring
  $RG$ and the  ring of
$RG$-matrices, see e.\ g.\ \cite{hur2}. 

Thus the methods of \sref{units1} and/or  \sref{pow2} may be
used to define convolutional codes  using group rings  $R=FG$ as 
 a subring  of $F_{n\times n}$ and then forming $R[z,z^{-1}] \cong
   RC_\infty$, which is the group ring over $C_\infty$ with coefficients
   from the group ring $R = FG$.

To obtain units in $R[z,z^{-1}]$ (which includes $R[z]$) 
we are lead to consider zero-divisors and
 units in $R=FG$. 

$R=FG$ is  a rich source of zero-divisors, and units, 
 and consequently $R[z,z^{-1}]$ is a
rich source of units. 
There are methods available for constructing units and zero-divisors
in $FG$. If $F$ is a field, every non-zero element of $FG$ is either
a unit or a zero-divisor.  
What is required are units in $R[z]$,
where $R = FG$, a group ring, and these can be obtained by the  use of   
zero-divisors and units in  $R$ as coefficients of the powers of $z$.   

In what follows bear in mind that in $R[z,z^{-1}]$ it is possible and
desirable that   $R$
has zero-divisors and  units, as 
when $R$ is a group ring.

\section{Convolution codes from group rings}
\label{sec:convolution}

Suppose then $\di\sum_{i=-m}^{n} \al_iz^i \ti
\di\sum_{j=-m}^{n}\be_jz^j = 1$ in the group ring $RC_\infty =
R[z,z^{-1}]$ with
$\al_i\in R$ and $C_\infty$ generated by $z$. 
By multiplying through by a power of $z$ this is then  
$\di\sum_{i=0}^n\al_iz^i \ti \di\sum_{j=-m}^n\be_jz^j = 1$.

 The  case with  $m=0$ gives polynomials over $z$.
Here we have $\di\sum_{i=0}^{n}\al_iz^i\ti \di\sum_{i=0}^t\be_jz^j =
1$ where $\al_n \not = 0, \be_t\not = 0$ and looking at the
coefficient of $z^0$ it is clear that we must also have $
 \al_0 \not = 0, \be_0 \not =0$. This can be considered an an equation
 in $RC_\infty$ with non-negative powers. Solutions may be used to
 construct convolutional codes. 

By looking at the highest and lowest coefficients we
then have that $\al_0\ti \be_0 = 1$ and $\al_n\ti \be_t= 0$. Thus in
particular $\al_0$
is a unit with inverse $\be_0$ and $\al_n, \be_t$ are zero divisors.

Solutions of  the general equation   
$\di\sum_{i=0}^n\al_iz^i \ti \di\sum_{j=-m}^n\be_jz^j = 1$ can also be
used to form 
convolutional codes and  
polynomial generator matrices may  be derived from these. 

\section{Examples} 

\subsection{A prototype example} Let $R = \Z_2C_4$. Then $\al_0 = a+a^2+a^3$
satisfies $\al_0^2 = 1$ and $\al_2 = a+a^3$ satisfies $\al_2^2= 0$. 

Thus $w = \al_0 + \al_1z + \al_2z^2$ in $RC_\infty$ satisfies
$w^2 = \al_0\al_0 + z(\al_0\al_1 + \al_1\al_0) +z^2(\al_0\al_2+ \al_1^2  +
\al_2\al_0) + z^3(\al_1\al_2 + \al_2\al_1) + z^4(\al_2\al_2) = 1 +
z^2\al_1^2$, since the $\al_i$ commute. 
Now require that $\al_1^2 =0$ and then $w^2 = 1$. 

In
particular letting  $\al_1=\al_2$ implies  that $w^2 = 1$. However,
just to be different, consider  $\al_1 = 1+a^2$ and then also $\al_1^2 =0$.
 
Now $\al_0$ corresponds to the matrix $\left(\begin{array}{rrrr} 0 &1 & 1& 1 \\
  1&0&1&1\\ 1&1&0&1\\ 1&1&1&0\end{array}\right)$, $\al_2$ 
  corresponds to the matrix $\left(\begin{array}{rrrr}
  0&1&0&1\\1&0&1&0\\ 0&1&0&1\\
  1&0&1&0  \end{array}\right)$ and $\al_1$ corresponds to the matrix 
$\left(\begin{array}{rrrr}
  1&0&1&0\\0&1&0&1\\ 1&0&1&0\\
  0&1&0&1  \end{array}\right)$. 

Take the first two rows of $w$ to generate a convolutional  code and then
the last two columns of $w$ is the control matrix of this code.

This gives the following generator matrix:

$$G = \left(\begin{array}{rrrr}0&1&1&1 \\ 1&0&1&1\end{array}\right) +
  \left(\begin{array}{rrrr}1&0&1&0\\0&1&0&1\end{array}\right) z
  +\left(\begin{array}{rrrr}0&1&0&1\\1&0&1&0\end{array}\right)z^2$$ 

The control matrix is: 

$$\left(\begin{array}{rrr}1  &
  1 \\ 1& 1 \\0&1\\1&0\end{array}\right) +
  \left(\begin{array}{rrr}1&0\\0&1\\1&0\\0&1 \end{array}\right)z +
  \left(\begin{array}{rrr}0&1\\1&0\\0&1\\1&0 \end{array}\right)z^2$$

The code has length $4$ and dimension $2$.
 It may be shown that the free  distance of this code is $6$. 

This can  be generalised.
\section{Convolutional codes from nilpotent 
  elements}\label{sec:char2}

The following two theorems are useful in constructing  new classes of
convolutional codes. 

\begin{theorem}\label{thm:ideal}  Let $R=FG$ be the group
ring of a group $G$ over a field $F$ with characteristic $2$. Suppose
$\al_i\in R$ commute. Let $w =
\di\sum_{i=0}^{n} \al_iz^i \in RC_\infty$. Then $w^2 = 1$ if and only
if $\al_0^2 = 1, \al_i^2 = 0, i>0$. 
\end{theorem}

\begin{proof} The proof of this is straight-forward and is omitted.

\end{proof}

The following is a generalisation of \thmref{ideal}; its proof is also
straight-forward and is omitted.

\begin{theorem}\label{thm:ideal1}  Let $R=FG$ be the group
ring of a group $G$ over a field $F$ with characteristic $2$. Suppose
$\al_i\in R$ commute. Let $w =
\di\sum_{i=0}^{n} \al_iz^i \in RC_\infty$. Then $w^2 = z^{2t}$ if and only
if $\al_i^2 = 0, i \not = t$ and  $\al_t^2 = 1$. 
\end{theorem}

To then construct convolutional codes proceed as follows. 
Find elements $\al_i$ with $\al_i^2=0$ and units $u$ with $u^2=1$ in the
group ring $R$. Then form units in $R[z]$ or $R[z,z^{-1}]$ using 
\thmref{ideal}  or
\thmref{ideal1}. From these units,
 convolutional codes are defined using the methods   described 
in \sref{units1} or \sref{pow2}.

\subsection{Examples 1}

Consider now $\al_0 = a+a^2+a^3$ and for $i> 0$ define $\al_i = a+a^3$
or $\al_i =0$ in the group ring
$R=\Z_2C_4$. Then $\al_0^2 = 1$ and $\al_i^2 = 0, i >0$. We could also
take $\al_i = 1 +a^2$. 

Define $w(z)= \di\sum_{i=0}^n \al_iz^i$ in $RC_\infty$. By
\thmref{ideal}, $w^2= 1$.

The matrix corresponding to $\al_0$ is $\left(\begin{array}{rrrr} 0 &1
  & 1&  1 \\
  1&0&1&1\\ 1&1&0&1\\ 1&1&1&0\end{array}\right)$ and the matrix
  corresponding to $\al_i, i \not = 0$ is 

$\left(\begin{array}{rrrr}
  0&1&0&1\\1&0&1&0\\ 0&1&0&1\\
  1&0&1&0  \end{array}\right)$ or else is the zero matrix.

Now specify that the first two rows of $w$ give  the generator matrix 
and from this it follows that 
the last two columns of $w$ is a control matrix. 

This gives the following generator matrix:

$$G = \left(\begin{array}{rrrr}0&1&1&1 \\ 1&0&1&1\end{array}\right) +
  \de_1\left(\begin{array}{rrrr}0&1&0&1\\1&0&1&0\end{array}\right) z
  +\de_2\left(\begin{array}{rrrr}0&1&0&1\\1&0&1&0\end{array}\right)z^2 +
  \ldots 
  +\de_n\left(\begin{array}{rrrr}0&1&0&1\\1&0&1&0\end{array}\right)z^n$$ 

where $\de_i = 1$ when $\al_i \not = 0$ and $\de_i = 0$ when $\al_i =
0$. 

The control matrix is:

$$H= \left(\begin{array}{rrr}1  &
  1 \\ 1& 1 \\0&1\\1&0\end{array}\right) +
  \de_1\left(\begin{array}{rrr}0&1\\1&0\\0&1\\1&0 \end{array}\right)z +
  \de_2\left(\begin{array}{rrr}0&1\\1&0\\0&1\\1&0
  \end{array}\right)z^2+\ldots +
  \de_n\left(\begin{array}{rrr}0&1\\1&0\\0&1\\1&0 \end{array}\right)z^n.$$ 

The code has length $4$ and dimension $2$. 
  The free distance is at least $6$ for any $n\geq 2$ and in many
  cases it will be larger. Polynomials used for generating cyclic
  linear codes suitably converted to polynomials in $R[z]$ prove
  particularly useful and amenable -- see for example \sref{2-1} below. 
   
\subsubsection{Particular Example}
The $(4,2)$ convolutional code with
  generator and check  matrices as follows has free distance $8$.

$$G= \left(\begin{array}{rrrr}0&1&1&1 \\ 1&0&1&1\end{array}\right) +
  \left(\begin{array}{rrrr}0&1&0&1\\1&0&1&0\end{array}\right) z
  +\left(\begin{array}{rrrr}0&1&0&1\\1&0&1&0\end{array}\right)z^3    
+\left(\begin{array}{rrrr}0&1&0&1\\1&0&1&0\end{array}\right)z^4
  $$

$$H= \left(\begin{array}{rrr}1  &
  1 \\ 1& 1 \\0&1\\1&0\end{array}\right) +
  \left(\begin{array}{rrr}0&1\\1&0\\0&1\\1&0 \end{array}\right)z +
  \left(\begin{array}{rrr}0&1\\1&0\\0&1\\1&0
  \end{array}\right)z^3+ \left(\begin{array}{rrr}0&1\\1&0\\0&1\\1&0
  \end{array}\right)z^4$$

\section{Direct products: Turbo-effect} 
Examples of convolutional codes formed
using $\al_i$ with $\al_i^2=0$ in $FG$ have been produced. 
Consider now $F(G\ti H)$ and
let $w = \be\ti \al_i$ for {\em any} $\be \in FH$. Then $w^2 =
\be^2\al_i^2 = 0$. This expands enormously the range of available
elements whose square is zero. Note
also that over a field of characteristic $2$ if $\al^2 =0 = \ga^2$
then $(\al+\ga)^2 = 0$. 

For example in $\Z_2C_2$  the element $ 1 + a$ was used where $C_2$
generated by $a$. Then in $\Z_2(G\ti C_2)$ consider $\al =
\be(1+a)$ for any $\be \in \Z_2G$. Then $\al^2 = 0$. 

A simple example of this is $\Z_2(C_2\ti C_2)$ where $\al = (1+a)b+
(1+b)a = a+b$. The matrix of $a+b$ is $\left(\begin{array}{cc} A & B
  \\ B & A\end{array}\right)$ where $A = \left(\begin{array}{cc} 0 &1
    \\ 1&0  \end{array}\right)$ and $B= \left(\begin{array}{cc}1&0 \\
    0&1\end{array}\right)$. In forming $(4,2)$ convolutional codes we
    would only use the top half of the matrices,
    i.e. $P= \left(\begin{array}{cc|cc}0&1 &1&0 \\ 1 &0&0&1
    \end{array}\right)$.
Note that in this encoding the vector $(\ga,\de)$ is mapped to
$(\ga,\de)P = \left(\begin{array}{cc|cc} \de & \ga & \ga &
  \de\end{array}\right)$. This
is like an interweaving of two codes.
 
To get a permutation effect, use the direct product with $S_n$, the
permutation or symmetric group on $n$ letters.

\section{(2,1) codes}\label{sec:2-1} See \cite{mac} for  examples of
$(2,1)$ optimal codes up to degree $10$. 
These can be reproduced algebraically and properties derived 
using the methods developed here. 

Further new $(2,1)$ convolutional codes
and series of convolutional $(2,1)$ are  constructed in this section 
 as an application of the 
general methods described above. 
The free distances can often  be determined
algebraically and codes to a prescibed free distance can be  
constructed by using \thmref{best} below. 

 Let $F$ be a field of
characteristic $2$ and $R=FC_2$, where $C_2$ is generated by $a$. 
Consider elements $\al_i\in R$, $i>0$, where either  $\al_i = 1+a$  or
$\al_i= 0$.  Then $\al_i^2 = 0$.

Let $\al_0 =1 $ in $R$ and define $w = \al_1 + \al_0z +
\al_2z^2+\ldots +\al_nz^n$. Then $w^2 = z^2$ and hence $w\ti (w/z^2) =
1$. Thus  $w$ 
can be used to define a $(2,1)$ convolutional code. 

More generally let $t$ be an integer, $0\leq t\leq n$, and 
define $w = \di\sum_{i=0}^n \be_iz^i$ where $\be_i =
 \al_i, i \not = t$, $\be_t = 1$. Then $w^2=
 z^{2t}$ gives that  $w\ti (w/z^{2t}) = 1$. Thus $w$ can be used to define a
 convolutional $(2,1)$ code. 
The case 
$\al_0= \be_1$ is a special case. 

Now determine the code by choosing the first row of the matrix of $w$ to
be the generator/encoder matrix and then the last column of $w/z^{2t}$ 
is the control matrix.

The matrix of $\al_i$ is $\left(\begin{array}{cc} 1& 1 \\
  1&1\end{array}\right)$ when $\al_i = 1+a$ and is the zero $2\ti 2$ 
matrix  when $\al_i = 0$. .


Define  $\de_i = 1 $ when $\al_i\not =0$ and  $i \not = t$;
$\de_i = 0$ when $\al_i = 0$ and  $i \not = t$; and define 
  $\de_t(1,1)$ to be $(1,0)$.

Then the encoder matrix of the code is 
$G = (1,1) + \de_1(1,1)z+ \de_2(1,1)z^2 + \ldots + \de_n(1,1)z^n$
and with 
$H= \left(\begin{array}{c} 1 \\ 1
\end{array}\right)+\de_1\left(\begin{array}{c} 1 \\ 0 \end{array}\right)z +
\de_2\left(\begin{array}{c} 1 \\ 1
\end{array}\right)z^2 + \ldots + \de_n\left(\begin{array}{c} 1 \\ 1
\end{array}\right)z^n$, the
control matrix is $H/z^{2t}$.

The generator matrix $G$ obtained in this way is noncatastrophic as it has a
right finite weight inverse -- see Theorems 6.3 and 6.6 in
\cite{mac}. 

For $n=2$ we get as an example the code with 
the generator matrix $G = (1,1)+(1,0)z+(1,1)z^2$. 
This code has free distance $5$ which is optimal. 
It is precisely the $(2,1,2,5)$ code
as described in \cite{mac}, page 1085.

\begin{theorem}\label{thm:five} $G$ has free distance $5$.
\end{theorem}
\begin{proof} Consider $\di\sum_{i=0}^t\be_iz^i G$, with $\be_i\in
  \Z_2$ and $\be_t \not = 0$. In determining free distance we may
  consider 
$\be_0 \not = 0$. The coefficients of $z^0 = 1$ and $z^{t+2}$ are
 $(1,1)$,  and also $(1,0)$ occurs in the expression for at least one
other coefficient. Thus the free distance is $2+2+1$ which is attained
by $G$.

\end{proof}

The above proof illustrates a general method for proving free distance
or getting a lower bound on the free distance. For example wherever 
$(1,0)$ appears in a sum making up a coefficient it will contribute a 
distance of at least $1$ as the other non-zero coefficients, all
$(1,1)$, will add up
to $(1,1)$ or $(0,0)$. 

The check matrix for this code is $\frac{\left(\begin{array}{c} 1\\ 1
\end{array}\right)+ \left(\begin{array}{c} 0\\ 1
\end{array}\right)z+ \left(\begin{array}{c} 1\\ 1
\end{array}\right)z^2}{z^2} = \left(\begin{array}{c} 1\\ 1
\end{array}\right)+ \left(\begin{array}{c} 0\\ 1
\end{array}\right)z^{-1}+ \left(\begin{array}{c} 1\\ 1
\end{array}\right)z^{-2}$.

For $n\geq 3$ it may be verified directly by similar algebraic methods
that the free
distance is at least $6$. Appropriate choices of the $\al_i$ will give
bigger free distances. See \thmref{best} below.

For $n=3, and \de_2=1=\de_3$  a $(2,1,3,6)$ convolutional code
is obtained which  is also optimal. Thus a 
degree $3$ optimal distance $6$ is given by 
 the encoder matrix  
$G = (1,1) + (1,0)z+ (1,1)z^2 + (1,1)z^3$
and the control matrix is   
$H/z^2= \left(\begin{array}{c} 1 \\ 1
\end{array}\right)/z^2+\left(\begin{array}{c} 1 \\ 0 \end{array}\right)/z +
\left(\begin{array}{c} 1 \\ 1 \end{array}\right) + 
\left(\begin{array}{c} 1 \\ 1 \end{array}\right)z$. It is clear that
$H$ is also a control matrix. 

The next case  is $(2,1,4)$ of degree $4$. The optimal distance of one
of these is
$7$.
Consider $w = \al_1 + \al_0z +\al_1z^3+\al_1z^4$, where $\al_1= 1+a$
and $\al_0 = 1$ in $\Z_2C_2$. Then $w^2 = z^2$ and thus $w$ gives the
encoder matrix and $w/z^2$ gives the check matrix.
The encoder matrix is $G= (1,1) + (1,0)z+ (1,1)z^3+(1,1)z^4$. Call
this code $\CC$. 

\begin{theorem}\label{thm:six} The free distance of $\CC$ is $7$.
\end{theorem}
\begin{proof} Consider $(\di\sum_{i=0}^t\be_iz^i) G$, with $\be_i\in
  \Z_2$. In determining free distance we may consider $\be_0 \not = 0$
and $\be_t \not = 0$. The coefficients of $z^0 (= 1)$ and $z^{t+4}$ are
both  $(1,1)$. If there are more than two non-zero $\be_i$ in the sum then
$(1,0)$ occurs in at least three coefficients giving a distance of
$2+2+3=7$ at least. It is now necessary to consider the case when there
are just two $\be_i$ in the sum. It is easy to see then that at least
three of the coefficients of $z^i$ are $(1,1)$, and $(1,0)$ or $(0,1)$
is a coefficient of another. Thus the free distance is $7$.
\end{proof}

\vspace{.05in}

Consider the next few degrees. Let $\al= 1+a, \al_0 = 1$ in $F C_2$
where $F$ has characteristic $2$.

\begin{enumerate}
\item deg 5: $w = \al+\al_0z + \al z^3+\al z^4+\al z^5$; gives a free
distance of $8$. 
\item 
deg 6: $w=\al + \al z^2+\al z^3+\al_0z^4+\al z^5+ \al z^6$. This
gives a free distance of $10$.

\item 
Consider for example the following degree $12$ element. 

$w = \al + \al z^2 +\al z^4+\al z^5+\al z^6+ \al_0z^9+
\al z^{10}+ \al z^{11} + \al z^{12}$

Note that this resembles the polynomial used for the Golay $(23,12)$
code -- see e.g. \cite{blahut} page 119. 
 The difference is that a $z^{12}$ has been added and the
coefficient of $z^9$ appears with coefficient $\al_0$ and not $0$ as
in the Golay code. It is possible to play around with this 
by placing $\al_0$ as
the coefficient of other powers of $z$ in $w$. 
\end{enumerate}

We thus study the best performance of convolutional codes derived from
$w = \di\sum_{i=0}^t\al_iz^i$ where
some $\al_t = 1 \in FC_2$, and all the other $\al_i$ are either $0$
or else $1+a$ in $ F C_2$. Try to choose the $\al_i$ as one would for a
linear cyclic code so as to maximise the (free) distance.

The set-up indicates we should look at existing cyclic codes and
form  convolutional codes by mimicking the generating polynomials
for the cyclic codes.

\subsection{From cyclic codes to convolutional codes}

Suppose now $\CC$ is a (linear) cyclic $(n,k,d_1)$ code over the field $F$ of
characteristic $2$. Suppose also  that  the dual of
$\CC$, denoted $\hat{\CC}$, is an $(n,n-k,d_2)$ code. 

Let $d = min(d_1,d_2)$. Suppose 
$f(g)= \di\sum_{i=0}^r\be_ig^i$, with $\be_i \in F, (\be_r\not = 0)$, 
is a generating polynomial for $\CC$. In $f(g)$, assume $\be_0 \not =
0$. 


Consider $f(z) = \di\sum_{i=1}^r \al_iz^i$ where now $\al_i = \be_i
\al$ with $\al = 1 + a$ in $FC_2$ or else $\al_i = 0$. 
Replace some $\al_i$, say $\al_t$, by $1$ or $a$  (considered as
members of $FC_2$).

So assume  $f(z) = \di\sum_{i=0}^r \al_iz^i$ with this
$\al_t = 1$  and other  $\al_i = \be_i\al$ so that
$\al_i = 1+a$ or $\al_i = 0$ (for $i \not = t$). It is also allowed to
let $\al_t = a$. 

Then  $f(z)^2 = z^{2t}$ and thus $f(z)\ti (f(z)/z^{2t}) = 1$. We now use 
$f(z)$ to generate a convolutional code by taking just the first rows 
of the $\al_i$. Thus the generating matrix is $\hat{f}=
\di\sum_{i=0}^r\hat{\al}_iz^i$ where  $\hat{\al}_i$ is the
first row of $\al_i$. 
\begin{lemma}\label{lem:once} Let $G$ be a generator matrix of a
  linear code  $\CC$ and
  suppose the dual code of $\CC$, $\hat{\CC}$,  has distance $d$. 
Then no row of $G$
  is a combination of less than $d-1$ other rows of $G$. 
\end{lemma}

\begin{proof} Now $G\T$ is the check matrix of $\hat{\CC}$. Since
  $\hat{\CC}$ has distance $d$ any $d-1$ columns of $G^T$ are linearly
  independent -- see e.g.\ \cite{blahut}, Corollary 3.2.3, page 52. 
Thus no column of $G\T$ is a combination of
  less than
  $d-1$ other columns of $G\T$.
 Hence no row of $G$ is a combination of less
  than $d-1$ other rows of $G$.

\end{proof} 

 \begin{lemma}\label{lem:twice}
 Let $w = \di\sum_{i=1}^n \al_i(1,1) + \al(1,0)$ 
 with $\al \not = 0$. Then
 at least one component of $w$ is not zero.
\end{lemma}

\begin{proof} Now $w = (\di\sum_{i=1}^n\al_i + \al,
 \di\sum_{i=1}^n\al_i) $. Since $\al \not = 0$ it is clear that one
 component of $w$ is not zero.
\end{proof}

 A similar result holds for  $w = \di\sum_{i=1}^n \al_i(1,1) +
 \al(0,1)$.

For the following theorem assume  the invertible element $\al_0$ does
not occur in the first or the last position of $f$; if it does occur
in one of these
positions, a similar result holds but the free distance is possibly
less by $1$.

\begin{theorem}\label{thm:best} Let $\CC$ denote the convolutional code
  with generator  matrix $\hat{f}$. Then the free distance of $\CC$ is
  at least $d+2$.
\end{theorem}

\begin{proof} 

 Consider  $ w = \di\sum _{i=0}^t\be_iz^i\hat{f}$ and we wish to show that
 its free distance is $\geq d+2$.
 In
calculating the free distance of $w$ we can assume $\be_0 \not =
0$ and we also naturally assume $\be_t \not = 0$. Let $fd(w)$ denote
 the free distance of $w$.

Let $w_1 = \di\sum_{i=0}^t\be_iz^i$. The support of $w_1$,
$supp(w_1)$, is the number of non-zero
$\be_i$. Suppose then 
$supp(w_1) \geq d$. Then in $w$, $\al_0$ appears with the coefficient
of  $z^i$,
  for at least $d$ different $i$ with $0 <
i < t+r$. Also the coefficient of $1 =
z^0$ is
$\be_0(1,1)$ and the coefficient of $z^{t+r}$  is $\be_t(1,1)$ and
each of these have distance $2$.
 Then by \lemref{twice}, $w$ has free distance at least
$d + 2$.

Consider $f(g) = \di\sum_{i=0}^r \be_ig^i$ and
$H(g)= f(g)(\di\sum_{i=0}^l\de_ig^i)$, with $l\leq k-1$ where $k$ is the rank
of the cyclic code. Then as this cyclic code has distance $d_1$,
$H(g)= \di\sum_{i=0}^{n-1}\ga_ig^i$ has support at least $d_1$.
Now $H(z) = f(z)(\di\sum_{i=0}^l\de_iz^i)$ is such that the sum of the
coefficients of $z^i, z^{i+n}, \ldots$ is $\ga_i$ for each $i$. Hence
if $\ga_i \not = 0$,
at least one of the coefficients of $z^i, z^{i+n}, \ldots $ is  not
$0$. Since $H(g)$ has support $d_1$, this implies that $H(z)$ has
support at least $d_1$. Hence $w$ has free distance at least 
 $(d_1-2) + 2\ti 2 = d_1 + 2 \geq d+2$ when $t\leq (k-1)$.

Assume then in $w$ that $t\geq k$. 
and that  $supp(w_1) < d $. If $supp(w_1) = 1$ then clearly $fd(w)
\geq (r-2) + 4= r +2 \geq d+2$.

Assume by induction that a sum such as $w$ of less than $t$ elements
with support less than $d$ has free distance at least $d+2$. 
     
Consider $f(g) = \di\sum_{i=0}^r \be_ig^i$ and
$H(g)= f(g)(\di\sum_{i=0}^l\de_ig^i)$, where $t > k-1$. 

Now as $\CC$ has $\rank k$, $f(g)g^k =
\di\sum_{i=0}^{k-1}\de_if(g)g^i$. Thus multiplying through by
$g^{t-k}$ implies $f(g)g^t =  \di\sum _{i=0}^{k-1}\de_ig^{i+t-k}f(g) =
\di\sum_{j=t-k}^{t-1}\de_{j-(t-k)}f(g)g^j$.

Now as $\hat{\CC}$ has distance $d_2$ the support of 
$\di\sum_{i=1}^{k-1}\de_if(g)g^i$ and hence of
$\di\sum_{j=t-k}^{t-1}\de_{j-(t-k)}f(g)g^j$  is at least $d_2-1$ by
\lemref{once}.

Now $\di\sum _{i=0}^{t-1}\be_iz^i\hat{f}$ has support at most $d-2$ as $w$
has support at most $d-1$.

Then $ w = \di\sum _{i=0}^t\be_iz^i\hat{f} = \di\sum
_{i=0}^{t-1}\be_iz^i\hat{f} +
\be_tz^t\hat{f} = \di\sum _{i=0}^{t-1}\be_iz^i\hat{f} + 
\be_t\di\sum_{j=t-k}^{t-1}\de_{j-(t-k)}\hat{f}
z^j = \di\sum_{i=0}^{t-1}\om_i\hat{f}z^i$ 
  and this sum is of non-zero support. Thus by induction the $fd(w) \geq
d+ 2$.



\end{proof}  

The free distance may be bigger than $d+2$;  an upper
bound is $2d-1$. The free distance also depends on where the
invertible $\al_0$ is placed in the expression for $f$. 
Placed near the `centre' will possibly give the best free distance.  

It is worth noting  
that if the support of the input element is $\geq
t$ then the free distance is at least $t+2$; this may be seen from the
proof of \thmref{best}. Thus it is possible to
avoid short distance codewords by ensuring that the input elements
have sufficient support -- this could be done by, for example, taking
the complement of any element with small support.  

The best choice for $\CC$ is probably a self-dual code as in this case
$d_1=d_2= d$.

There exist self-dual codes of arbitrary large distances. See also
\cite{hur3} for many constructions of self-dual codes.

These convolutional codes can be considered to be self-dual type convolutional 
codes in the sense that
$f(z)$ determines the generator matrix  and $f(z)/z^{2t}$ determines the 
control matrix.

\section{(2m,1) codes} The previous section \sref{2-1} can be
 generalised to produce convolutional codes of smaller rate $(2m,1)$
 but with much bigger free distance. Essentially the free distance is
 multiplied by $m$ over that obtained for similar $(2,1)$ codes.

The group to consider is $C_{2m}$ generated by $a$. Assume $m$ is odd
although similar results may be obtained when $m$ is even. 
 Let $\al = 1+a+a^2+
\ldots + a^{2m-1}$ and $\al_0 = 1+ a^2 + \ldots + a^{2m-2}$. Then
$\al^2 = 0$ and $\al_0^2 = 1$ as $\al_0$ has odd support. 

Define as before $f(z) = \di\sum_{i=1}^r \al_iz^i$ where now $\al_i = \be_i
\al$ in $\Z_2C_{2m}$ or else $\al_i = 0$. 
Replace some $\al_i$, say $\al_t$, by $\al_0$.

Then $f(z)^2 = z^{2t}$ and $f(z)(f(z)/z^{2t}) = 1$. Thus use $f(z)$ to
define a convolutional code $\C$ by taking the first row of the $\al_i$.

For example $G(z) = (1,1,1,1,1,1) + (1,0,1,0,1,0)z + (1,1,1,1,1,1)z^2$
defines a $(6,1)$ convolutional code which has free distance $15$.  
$G(z) = (1,1,1,1,1,1) + (1,0,1,0,1,0)z + (1,1,1,1,1,1)z^3+
(1,1,1,1,1,1)z^4$ defines a convolutional code which has free distance
$21$.

A theorem similar to \thmref{best} is also true:
Let $f(g)$ denote the generator matrix of a cyclic code with distance
$d_1$ and whose dual code has distance $d_2$. Let $d = \min(d_1,d_2)$
and let $\CC$ denote the convolutional code obtained from $f(z)$ where
the coefficients of $f(g)$ have been replaced by $\al_i$ in all but one
coefficient which has been replaced by $\al_0$ and the first row of
each coefficient is used. Assume in the following theorem that $\al_0$ is not
in first or last coefficient. 
\begin{theorem}
The free distance of $\CC$ is at least $md+2m$.
\end{theorem}

\section{Higher rates}\label{sec:higher} The methods of \sref{2-1} can also be generalised to
 produce higher rate convolutional codes.

Consider achieving a rate of $3/4$.

In $C_4$ generated by $a$, define $\al= 1+ a$ and $\al_0 = 1$. Then
$\al^4= 0$ and $\al$ can be used to define a code of rate $3/4$ and
distance $2$. Now $\al$ has matrix

$\left(\begin{array}{cccc} 1 &1&0&0\\ 0 &1&1&0\\ 0&0&1&1 \\ 1&0&0&1  \end{array}\right)$

and the first three rows of this 

$A= \left(\begin{array}{cccc} 1 &1&0&0\\ 0 &1&1&0\\ 0&0&1&1 \end{array}\right)$
 
generates a $(4,3,2)$ code.

Now the matrix of $\al_0$ is $I_{4\ti 4}$, the identity $4 \ti 4$
matrix and let $B$ denote the first three rows of $I_{4\ti 4}$.

\begin{lemma}\label{lem:vec} Let $\un{x} \not = 0 $ be  a $1\ti 3$ vector. 
Then $\un{x} (A +  B)$  is  not the 
  zero vector and  thus $\un{x} (A +  B)$  has distance at least $1$. 
\end{lemma}

\begin{proof} Now $(\al+ 1)^4 = \al^4+ 1 = 1$ and so $(\al+1)$ is a
  non-singular matrix. Thus in particular the first three rows of the
  matrix of $(\al+1)$ are linearly independent. The first three rows
  of $\al + 1$ precisely constitutes the matrix $A+B$. Thus $\un{x}
  (A+B)$ is not the zero vector. 

Another way to look at this is that $\al+1 = a$ but it is useful to
look at the  more general way in \lemref{vec} for further developments.

\end{proof}

\begin{corollary}
If $\un{x} A + \un{y}B = \un{0}$ then $\un{x} \not = \un{y}$.
\end{corollary}

Form convolutional $(4,3)$ codes as follows.

Let $f(z) = \di\sum_{i=0}^n \al_i z^i$ where $\al_i = \al$ or $\al_i=
0$ except for $\al_t = 1$ for some $t, 1 < t \leq n$. We could also
use $\al_1 = \al_t = 1$ but this generally gives smaller distance codes.
 
Then $f(z)^4 = z^{4t}$ and so $f(z)\ti (f(z)^3/z^{4t}) =1$. Thus use
$f(z)$ to generate the code and  $(f(z)^3/z^{4t})$ to check/control
the code. Take the first
three rows of the matrix of $f(z)$ to generate a $(4,3)$ code and
delete the last three columns $(f(z)^3/z^{4t})$ to form the control
matrix. 

Thus $G(z) = \di\sum_{i=0}^n \hat{\al_i} z^i$ is the generator matrix where
 $\hat{\al_i}$ is the first three rows of the matrix
of $\al_i$. 

In \sref{2-1} we had the situation that when $\al_0$ occurred in any
coefficient then it
contributed a distance of $1$, so that when the support of $G$ is $s$ then
$\al_0$ will contribute a free distance of $s$. Here we us the fact
that if $\al_0$ occurs then it will contribute a distance of at least
$1$ unless its coefficient equals the sum of the coefficients in the
other non-zero $\al_i$ which occur with it in the same coefficient of
$z^j$.  
       
\subsection{Examples} The generator
matrix 
$$G = \left(\begin{array}{cccc} 1 &1&0&0\\ 0 &1&1&0\\ 0&0&1&1
\end{array}\right) + \left(\begin{array}{cccc} 1 &0&0&0\\ 0 &1&0&0\\
  0&0&1&0  \end{array}\right)z+
 \left(\begin{array}{cccc} 1 &1&0&0\\ 0 &1&1&0\\
  0&0&1&1  \end{array}\right)z^2$$
 
defines a $(4,3)$ convolutional code. It may be shown that its free
distance is $5$. The proof is similar to the proof of \thmref{five}
but also using \lemref{vec}.

The check matrix for the  code is easy to write out. 

Consider  $n=3$, and   
$$G = \left(\begin{array}{cccc} 1 &1&0&0\\ 0 &1&1&0\\ 0&0&1&1
 \end{array}\right) +\left(\begin{array}{cccc} 1 &0&0&0\\ 0 &1&0&0\\
   0&0&1&0 \end{array}\right)z +
 \left(\begin{array}{cccc} 1 &1&0&0\\ 0 &1&1&0\\ 0&0&1&1
 \end{array}\right)z^2 +
\left(\begin{array}{cccc} 1 &1&0&0\\ 0 &1&1&0\\ 0&0&1&1
 \end{array}\right)z^3 $$
 
This is a $(4,3)$ convolutional code and its free distance is  $6$.  

The next example  is 
$$G = \left(\begin{array}{cccc} 1 &1&0&0\\ 0 &1&1&0\\ 0&0&1&1
 \end{array}\right) +\left(\begin{array}{cccc} 1 &0&0&0\\ 0 &1&0&0\\
   0&0&1& 0\end{array}\right)z +
 \left(\begin{array}{cccc} 1 &1&0&0\\ 0 &1&1&0\\ 0&0&1&1
 \end{array}\right)z^3 +
\left(\begin{array}{cccc} 1 &1&0&0\\ 0 &1&1&0\\ 0&0&1&1
 \end{array}\right)z^4 $$
 
This has free distance $7$. This may be proved similar to
\thmref{six} using \lemref{vec}.

It is then possible to proceed as in \sref{2-1} to investigate further
degrees (memories) with rate $3/4$.





\subsection{Polynomial} In cases where a polynomial
 generator {\em and} polynomial right
 inverse for this generator  are required, insist that $\al_0= 1$. This gives
 slightly less free distance but is interesting in itself.  





%
For example consider  the encoder matrix  
$G = (1,0) + \de_1(1,1)z+ \ldots + \de_n(1,1)z^n$
and the control matrix is 
$H= \left(\begin{array}{c} 1 \\ 0
\end{array}\right)+\de_1\left(\begin{array}{c} 1 \\ 1 \end{array}\right)z +
\ldots + \de_n\left(\begin{array}{c} 1 \\ 1
\end{array}\right)z^n$. Here $\de_i = 0$ or $\de_i =1$. 

This code has free distance $4$ for $n = 2$. For $n\geq 2$ the free
distance will depend on the choice of the $\de_i$. As already noted,
the choices where the $z$-polynomial corresponds to a known cyclic code
polynomial deserves particular attention.

  
We may also increase the size of the field as for example as follows. 

Consider now $R = GF(4)C_2$, the group ring of the cyclic group of
order $2$ over the field of 4 elements. Define $\al_0 = \om + \om^2g$,
$\al_1= \om + \om g, \al_2= \om^2+\om^2g$,  where $\om$ is the primitive
element in $GF(4)$ which satisfies $\om^2 + \om + 1 = 0,\om^3 = 1$. 
Then $\al_0^2 = \om^2+ \om^4 = \om^2 +\om = 1$ and $\al_1^2 = \al_2^2
= 0$.  
Thus  
$w = \al_0 + \al_1z +
\al_2z^2$ satisfies $w^2 = 1$ and can be used to
define a convolutional code of length $2$ and dimension $1$. 
The encoder matrix is then 
 $G = (\om,\om^2) + \de_1(\om,\om)z+  \de_2(\om^2,\om^2)z^2 + ...+
\de_n(\om^i,\om^i)z^n$
and the control matrix is 
$H= \left(\begin{array}{l} \om^2 \\ \om
\end{array}\right)+\de_1\left(\begin{array}{c} \om  \\ \om
\end{array}\right)z  +
\ldots + \de_n\left(\begin{array}{c} \om^i \\ \om^i \end{array}\right)z^n$.

 The {\em degree} of a convolutional code with encoder matrix $G(z)$ is
  defined to be  the maximal degree of the full $k\ti k$
  size minors of $G(z)$ where $k$ is the dimension;  see
  \cite{blahut}. The maximum free
  distance of a length 2, dimension one, degree $\delta$ code over any
  field is by
  \cite{ros}, $2\delta + 2$.  
 
Consider the case $n=2$. The encoder matrix is then 
 $G = (\om,\om^2) + (\om,\om)z+  (\om^2,\om^2)z^2$.
The degree of this
code is $\delta = 2$ since the dimension is $1$. 
Let $G' = (1,\om) + (1,1)z + (\om,\om)z^2$ so that $\om G' = G$. 
\begin{theorem}\label{thm:freed} The free distance of this code  is
 $6$ and  so is thus  
a maximum distance separable convolutional code.
\end{theorem}
\begin{proof} Consider combinations $(\al_0 + \al_1z+ \ldots +
  \al_tz^t)G$ and we wish to show that this has (free) distance
  $6$. 
We may
  assume $\al_0\not = 0$. It is clear  when $t=0$ that $w$ has a
  distance of $6$ and so in particular a distance of $6$ is attained.  
Since also $\om$ is a factor of $G$ we
  may now consider the minimum distance of 
$w = (\al_0+\al_1z+ \ldots + \al_tz^t)G'$ with $\al_0\not =
  0, \al_t\not = 0$ and $t>0$.  
The coefficient of $z^0$ is $\al_0(1,\om)$; the coefficient of
  $z^{t+2}$ is $\al_t(\om,\om)$, the coefficient of $z^{t+1}$ is
  $\al_t(1,1) + \al_{t-1}(\om,\om)$ and the coefficient of $z^t$ is
  $\al_t(1,\om) + \al_{t-1}(1,1) + \al_{t-2}(\om,\om)$ when $t\geq 2$ 
 and the coefficient of $z$ is  $\al_1(1,\om)+\al_0(1,1)$ and 
 this is also the case when $t=1$.  

Case $t\geq 2$: If $\al_t\not = \al_{t-1}\om$ then the coefficient of
  $z^{t+1}$ has distance $2$ giving a distance of $6$ with $2$ coming
  from each of the coefficients of $z^0,z^{t+1},z^{t+2}$. If $\al_t
  = \al_{t-1}\om  $ the coefficient of $z^{t}$ is
  $\al_{t-1}(\om+1,\om^2+1) + \al_{t-1}(\om,\om)$; in any case this
  has distance $\geq 1$. Also the coefficient of $z$ has distance
  $\geq 1$. Thus the total distance is at least $2+1+1+2=6$.

Case $t=1$. If $\al_0 \om \not = \al_1$ then the coefficient of $z^2$
  has distance $2$ and thus get a distance of $2+2+2=6$ for the
  coefficients of $z^0,z^2,z^3$. If $\al_0 \om = \al_1$ then the
  coefficient of $z$ is $\al_1(1,\om) + \al_0(1,1) = \al_0(\om+1, \om^2
  +1)$ which has distance $2$. Thus also we get a distance of $2+2+2 =
  6$ from coefficients of $z^0,z,z^3$.   

Note that the proof depends on the fact that
  $\{1,\om\}$ is linearly independent in $GF(4)$.
\end{proof}


\subsubsection{Bigger fields}
It will be necessary to work over bigger fields to get length 2,
dimension $1$,  maximal distance
separable convolutional codes of higher degree. 

Consider $\F = GF(2^n)$ with generating  element $\om$ satisfying
$\om^n+\om+1 = 0$. Then $w_0 = \om + \om^na$ in $\F C_2$, where $C_2$
is generated by $a$ 
satisfies $w_0^2 = \om^2+\om^{2n} = 1$ since $\om^n = \om + 1$ and 
$w_i= \om^i+\om^i$, defined for $i> 0$, satisfies $w_i^2 = 0$. 

A generating element is then formed from these $w_i$. 
Consider $w(z) = w_0 + \de_1w_{i_1}z+ \ldots + \de_nw_{i_n}z^n$ where  
$w_{i_j}$ is some $w_i$ and $\de_i \in \{0.1\}$.  
Then  $w(z)^2 =1$ and is then 
used to define a convolutional code of length $2$ and dimension
$1$. 

The $w_0$ can be taken as the coefficient of any $z^t$ in the
definition of $w(z)$ and
convolutional codes are similarly defined.

The further study of these codes is not included here.




\section{General rank considerations}


Let  
$w(z) = \di\sum_{i=0}^t \al_iz^i$ where $\al_i^2 = 0, i\not = t, \al_t^2 =
1$ with the $\al_i$ in some group ring $RG$. Suppose the $\al_i$
commute and that $R$ has characteristic $2$. Then 
$w(z)^2 = z^{2t}$. 

Consider the ranks of the non-zero $\al_i$ in 
deciding which rows of  $w$ to choose with which to construct the
convolutional code. For example if 
the non-zero  $\al_i$ satisfy 
$\rank \al_i = 1/2|G| = m$ we choose the matrix with just half
the rows of the matrix of each $\al_i$.
 

Many good codes may be produced this way.

It is possible to have more than one $\al_t$
satisfying $\al_t^2=1$ in $w(z)$ but then  the 
generator matrix  produced can be  catastrophic, although a valid code
may still be defined.

\subsection{Example} Let $u = 1 +h(a+a^2+a^3)$ in $\Z_2(C_4\ti C_2)$.
Then $u^2=0$ and $\rank u =4$. Define $w = u + z + uz^2$. Then $w^2 =
z^2$ and $w$ is used to define a $(8,4)$ convolutional code. The
generator matrix is 
$G = (I,B) + (I,0)z+(I,B)z^2$ where $B =
\left(\begin{array}{cccc}0&1&1&1\\1&0&1&1\\
  1&1&0&1\\1&1&1&0\end{array}\right)$. Now $(I,B)$ has distance
  $4$. Any combination of $(I,B), (I,0)$ has distance $1$ at least as
  $B$ is non-singular. Thus consider $(\di\sum_{i=0}^t\be_iz^t)G$. The
  highest and lowest power of $z$ has distance $4$ and there is a
  power of $z$ in between which has distance $1$ so altogether we get
  a free distance of $9$.  The degree of the code is $8$.

This can be extended. It can also be extended by finding higher
dimensional $u$ with $u^2=0$. See \sref{hamm} for further development
of these ideas. 

\subsection{Higher rates with nilpotent elements}\label{sec:higher1}
So far we have used $\al_i$ with $\al_i^2 =0$ and this generally give
rate 1/2 convolutional codes. We now look at elements $\al$ with
$\al^4=0$ with which to produce convolutional rate 3/4 codes. 
See \cite{hur3} for
 where such elements 
 are used to produce dual-containing codes.

See \sref{higher} for some preliminary examples on these. 

Suppose then $w = \di\sum_{i=0}^n\al_iz^i$ in $\F G$ 
where $\al_i^4=0, i\not = t
$ and 
$\al_t^4 =1, 1\leq t \leq n$. Suppose also $\F$ has characteristic $2$
and that the $\al_i$ commute. Then $w^4 = z^{4t}$. Thus
$w$ is used to generate a 3/4 rate convolutional code by taking
 the first 3/4 of the rows of the $\al_i$; then $w^3/z^{4t}$
will be the control matrix using the last 1/4 of the columns of the
$\al_i$.

For examples of elements $\al_i$ with  $\al_i^4=0$, see \cite{hur3}.

\subsubsection{Example} Consider $\al = a+a^7 \in \Z_2C_8$. Then
$\al_i^4 = 0$ and $\al$ generates an $(8,6,2)$ linear cyclic
code -- this is the best distance for a linear $(8,6)$ code. 
Now construct convolutional codes similar to the construction of
the $(2,1)$ codes. 

An element $\al_0 \in \Z_2C_8$ such that $\al_0^4 = 1$ is needed. There
are a number of choices including $\al_0 =1, \al_0 = 1+a+a^3, \al_0 =
1+ a+ a^7$. Choose $\al_0$ so that the first 3 rows of the matrix of
$\al_0$ generates a linear code of largest distance.  It is easy to
verify that the first three rows of $\al_0 = 1+a+a^3$ generates a
linear code of distance $2$.   

\begin{itemize}
\item $w = \al + \al_0z$. This gives a $(8,6)$ code of free distance
$4$. The `degree' in the convolutional sense  is $6$.

\item $w = \al + \al_0z+\al z^2$. This is  a $(8,6)$ convolutional 
 code of free distance
$6$. The `degree' here is $12$.

\item 
$w = \al + \al_0z+\al z^2+\al z^3$ gives an $(8,6)$ code of free
distance $6$.  

\item $w = \al + \al_0z +\al z^3+\al z^4$ gives an $(8,6)$ code of free
distance $8$. 
\item 

Polynomial degree 5: $w = \al +\al_0z + \al z^3+\al z^4+\al z^5$. The
free distance has to be determined.
\item Polynomial degree 6: $w=\al + \al z^2+\al z^3+\al_0z^4+\al z^5+
  \al z^6$. This should give  a free distance of at least $10$.
\item As for the $(2,1)$ convolutional 
codes in \sref{2-1}, by mimicking the polynomials
  used to generate cyclic codes, it
  should be possible to get $(8,6)$ convolutional codes with
  increasing free distance.  
\end{itemize}


\newcommand{\cc}{\mathbb{C}}
\section{Using idempotents to generate convolutional codes}
Let $FG$ be the group ring over a field $F$. For most cases in
applications it is 
required  that $\text{char}\, F 
\not | \, |G|$.  It  may also be necessary to  require that $F$
contains a primitive
$n^{th}$ root of unity.
 The complex numbers $F=\cc$ satisfies these conditions. 
 
 The reader is (again) referred to  \cite{seh} for background
 definitions and results on  group
 rings in relation to this section.

Let $\{e_1, e_2, \ldots, e_k\}$ be a complete family  of orthogonal
idempotents in $FG$. Such sets always exist when $\text{char} \, F \not | \, |G|$.

Thus:\\
(i) $e_i \not = 0$ and $e_i^2 = e_i$, $1\leq i\leq k$.\\ (ii) If
$i\not = j$ then $e_ie_j = 0$. \\ (iii) $1 = e_1+e_2 + \ldots + e_k$.

Here $1$ is used for the identity of $FG$.

\begin{theorem}\label{thm:first} Let $f(z) = \di\sum_{i=0}^k\pm
  e_iz^{t_i}$. Then
  $f(z){f}(z^{-1})=1$.
\end{theorem}
\begin{proof} Since $e_1, e_2, \ldots, e_k$ is a set of orthogonal primitive
idempotents, $f(z)f(z^{-1}) = e_1^2 + e_2^2+ \ldots + e_k^2 = 1$.

\end{proof}


The result in \thmref{first} can be considered as an identity in
$RC_\infty$ wherein $R=FG$ is a group ring. 

To now construct convolutional codes, decide on the rank $r$ and then
use the first $r$ rows of the matrices of the $e_i$ in
\thmref{first}. The control matrix is obtained from $f(z^{-1})$ by
deleting the last $r$ columns of the $e_i$.

If the $e_i$ have $\rank \geq k$ and for some $i$ $\rank e_i = k$ 
then it is probably best to take the $r = k$ for the rank of the
convolutional code, although other cases also have uses depending on
the application in mind.

\subsection{Idempotents in group rings}  Orthogonal sets of  
idempotents may be obtained in group rings from the conjugacy classes and 
character tables, see e.g.\ \cite{seh}. 

Notice also that a product $h(z)= \prod_if_i(z)$ where the $f_i(z)$ satisfy
 the conditions of \thmref{first} also satisfies $h(z)\hat{h}(z^{-1})
 = 1$, where $\hat{h}(z^{-1})$ is the product of the $f_i(z^{-1})$ in reverse
 order, 
 and thus $h(z)$ can then be used to define convolutional codes.  

In the ring of matrices define $e_{ii}$ to be the matrix with $1$ in the
$i^{th}$ diagonal and zeros elsewhere. Then $e_{11}, e_{22}, \ldots,
e_{nn}$ is a complete set of orthogonal idempotents and can be used to
define such $f(z)$. These  in a sense are trivial but can be useful
and can also be combined with  others.

To construct convolutional codes:
\begin{itemize}
\item Find
sets of orthogonal idempotents.
\item  Decide on the $f(z)$ to be used with each set.
\item Take  the product of the $f(z)$.
\item  Decide on the rate. 
\item Convert these idempotents into matrices as per the isomorphism 
between the group ring and a ring of matrices.
\end{itemize}

Group rings are a rich source of complete sets of orthogonal
idempotents.
This brings us into character theory in group rings. Orthogonal sets over the
rationals and other fields are also obtainable.

 

The Computer Algebra packages GAP and Magma can construct character tables
and conjugacy classes from which complete sets of orthogonal
idempotents may be obtained.

\subsection{Example 1} Consider $\C C_2$ where $C_2$ is generated by
$a$. Define $e_1 = \frac{1}{2}(1+a)$ and $e_2 =
1-e_1 = \frac{1}{2}(1-a)$. This gives $f(z) = e_1 +e_2z^{t}$ or $f(z)
= e_2 + e_1z^{t}$ for various $t$. Products of these could also be
used but in this case we get another of the same form by a power of $z$. 

\subsection{Cyclic} The orthogonal idempotents and character table of the
cyclic group are well-known and are closely related to the Fourier
matrix. 

This gives for example in $C_4$, $e_1 = \frac{1}{4}(1+a + a^2 +a^3),
e_2 =\frac{1}{4}( 1 + \om a+\om^2 a^2+\om^3 a^3), e_3 = \frac{1}{4}(1
- a + a^2  -a^3),
e_4 = \frac{1}{4}(1+\om^3a+\om^2 a^2 + \om a^3)$ from which $4\ti 4$
 matrices with degree $4$ in $z$ may be constructed, where $\om$ is a
 primitive $4^{th}$ root of unity. Notice
in this case that $\om^2 = -1$.  

Let $f(z) = e_1+e_2z+e_3+e_4z^3$. Then $f(z)f(z^{-1})= 1$. We take the
first row of the matrices to give the following generator matrix for a
$(4,1,3)$ convolutional code:

$G(z) = \frac{1}{4}\{(1,1,1,1)+(1,\om,-1,-\om)z + (1,-1,1,-1)z^2+
(1,-\om,-1,\om)z^3\}$.

It is easy to check that a combination of any one, two or three of the
vectors \\$(1,1,1,1),(1,\om,-1,-\om),(1,-1,1,-1),(1,-\om,-1,\om)$, which
are the rows of the Fourier matrix, has distance at least 2 and
a combination of all four of them has distance 1. From this it is easy
to show that the code has free distance $14$ -- any combination of
more than one will have 4 at each end and three in the middle with
distance at least $2$. This gives a $(4,1,3,14)$ convolutional codes
which is optimal -- see \cite{ros}. 

We can combine the $e_i$ to get real sets of orthogonal
idempotents. Note that it is enough to combine the conjugacy
classes of $g$ and $g^{-1}$ in order to get real sets of orthogonal
idempotents. 

In this case then we get 

$\hat{e}_1 = e_1 = \frac{1}{4}(1+a+a^2+a^3), \hat{e}_2 = e_2+e_4 = 
\frac{1}{2}(1 -a^2), \hat{e}_3 = e_3 = \frac{1}{4}(1-a+a^2-a^3)$, which can
then be used to construct real convolutional codes. 
 
Then $G(z) = \frac{1}{4}\{(1,1,1,1)+2(2,0,-2,0)z + (1,-1,1,-1)z^2\}$
gives a $(4,1,2)$ convolutional code. Its free distance is $10$ which
is also optimal.  

Using $C_2\ti C_2$ gives different  matrices. Here the set
of orthogonal idempotents consists of 
 $e_1 = \frac{1}{4}(1+a+b+ab), e_2 = \frac{1}{4}(1- a +b-ab), e_3=
 \frac{1}{4}(1-a-b +ab), e_4 = \frac{1}{4}(1+a- b-ab)$
and the matrices derived are all real. 
 
This gives $G(z) = \frac{1}{4}\{(1,1,1,1) + (1,-1,1,-1)z +
(1,-1,-1,1)z^2+(1,1,-1,-1)z^3\}$. Its free distance also seems to be
$14$. 

\subsection{Symmetric group} The orthogonal 
idempotents of the symmetric group are well-understood and are real.

We present an example here from $S_3$, the symmetric group on 3 letters.

Now $S_3 = \{1, (1,2), (1,3), (2,3), (1,2,3), (1,3,2)\}$ where these are
cycles. We also use this listing of $S_3$ when constructing matrices. 

There are three conjugacy classes:
$K_1 = \{1\}$; $K_2 = \{(1,2), (1,3),  (2,3)\}$; $K_3 =\{
(1,2,3),(1,3,2)\}$.

Define \\ $\hat{e}_1 = 1 + (1,2)+(1,3)+(2,3)+(1,2,3)+(1,3,2)$,
\\ $\hat{e}_2 = 1 - \{ (1,2) + (1,3)+(2,3)\} + (1,2,3)+(1,3,2)$,\\
$\hat{e}_3 = 2 - \{(1,2,3) + (1,3,2)\}$,

and $e_1 = \frac{1}{6}\hat{e}_1; e_2 = \frac{1}{6}\hat{e}_2; e_3 =
\frac{1}{3}\hat{e}_3$. Then $\{e_1,e_2,e_3\}$ form a complete
  orthogonal set of  idempotents and may be used to construct
  convolutional codes. 

The $G$-matrix of $S_3$ (see \cite{hur2}) is 

$$\left(\begin{array}{rrrrrr}1 & (12)&(13)&(23)&(123)&(132)\\
  (12)&1&(132)&(123)&(23)&(13) \\ (13) &(123) &1 & (132) &(12)&(23) \\
  (23) &(132)&(123)&1&(13)&(12) \\ (132)& (23)&(12)&(13)&1&(123) \\
  (123)&(13)&(23)&(21)&(132)&1\end{array}\right).$$

Thus the matrices of $e_1,e_2,e_3$ are respectively 

$E_1=\frac{1}{6} \left(\begin{array}{rrrrrr} 1 & 1&1&1&1&1\\
  1&1&1&1&1&1 \\ 1&1&1&1&1&1 \\ 1&1&1&1&1&1 \\ 1&1&1&1&1&1
  \\ 1& 1&1&1&1&1 \end{array}\right)$
 
$E_2=\frac{1}{6} \left(\begin{array}{rrrrrr} 1 & -1&-1&-1&1&1\\
  -1&1&1&1&-1&-1 \\ -1&1&1&1&-1&-1 \\ -1&1&1&1&-1&-1 \\ 1&-1&-1&-1&1&1
  \\ 1& -1&-1&-1&1&1 \end{array}\right)$

$E_3=\frac{1}{3} \left(\begin{array}{rrrrrr} 2 & 0&0&0&-1&-1\\
  0&2&-1&-1&0&0 \\ 0&-1&2&-1&0&0 \\ 0&-1&-1&2&0&0 \\ -1&0&0&0&2&-1
  \\ -1& 0&0&0&-1&2 \end{array}\right)$

Note that $e_1,e_2$ have $\rank 1$ and that $e_3$ has $\rank 2$. 





\section{Other characteristics}  Convolutional codes over fields of
arbitrary characteristic, and not just characteristic $2$, may also be 
constructed using the general method as previously described.

The following theorem is similar to \thmref{ideal1}. 

\begin{theorem}\label{thm:ideal111}  Let $R=FG$ be the group
ring of a group $G$ over a field $F$ with characteristic $p$. Suppose
$\al_i\in R$ commute and $\ga_i \in F$. Let $w =
\di\sum_{i=0}^{n} \al_i\ga_iz^i \in RC_\infty$. Then $w^p = \ga_t^pz^{pt}$
if and only
if $\al_i^p = 0, i \not = t$ and  $\al_t^p = 1$. 
\end{theorem}

The situation with $\ga_t = 1$ is easiest to deal with and is not a
great restriction. 
 
Construct convolutional codes  as follows. 
Find elements $\al_i$ with $\al_i^p=0$ and units $u$ with $u^p=1$ in the
group ring $R$. Then define  elements  as in \thmref{ideal111} in $R[z]$
to form units in $R[z]$. Thus get $f(z)^p = \ga_t^pz^{pt}$ and hence $f(z)\ti
f(z)^{p-1}/(\ga_t^pz^{pt}) = 1$. From these units,
convolutional codes are defined as  described 
in \sref{units} or \sref{pow2}.  
 
Thus
$f(z)$ may be used to define a convolutional code. By choosing the first
$r$ rows of the $\al_i$ considered as matrices defines a
$(n,r)$ convolutional code where $n = |G|$. The generator matrix is 
$\hat{f}(z) =
\di\sum_{i=0}^n \hat{\al_i}\ga_iz^i$ where $\hat{\al_i}$ denotes the first
  $r$ rows of the matrix of $\al_i$.  

It is necessary to decide which rows of the matrix to choose in
defining the convolutional code. This is usually decided by
considering the rank(s) of the non-zero $\al_i$. 

\subsection{Examples for characteristic 3}  Suppose then
$F$ has characteristic $3$ and consider $F(C_3\ti C_3)$ where the $C_3$
are generated respectively by $g, h$. 

Define $\al = 1 + h(1+g)$. Then $\al^3 = 0$. Define $\al_0 =
2+2h$. Then $\al_0^3 = 1$. 

The matrix of $\al $ is $P= \left(\begin{array}{rrr} I & B & 0 \\ 0 &I & B
\\ B&0&I \end{array}\right)$ where $I$ is the identity $3\ti 3$
matrix, $0$ is the zero
$3\ti 3$ matrix and $ B = \left(\begin{array}{rrr} 1&1 &0 \\ 0&1&1 \\ 1&0&1
\end{array}\right)$. 

By row (block) operations $P$ is equivalent to $\left(\begin{array}{rrr} I
  & 0 & -B^2 \\ 0 & I & B \\ 0&0&0 \end{array}\right)$. Thus $P$ has
  rank $6$ and the matrix $Q = \left(\begin{array}{rrr} I
  & 0 & -B^2 \\ 0 & I & B \end{array}\right)$ defines a block $(9,6)$
  code which indeed has distance $3$. 

Now define $\al_t = \al_0$ for some $0<t < n$ and 
choose $\al_i = 0$ or $\al_i = \al$ for $i\not = t$.
Define $f(z) = \di\sum_{i=0}^n\al_i z^i$. 
Then by \thmref{ideal111},
$f(z)^3 = z^{3t}$ and hence $f(z)\ti (f(z)^{2}/z^{3t}) = 1$. Thus
$f(z)$ may be used to define a convolutional code. Choose the first
$6$ rows of the $\al_i$ in $f(z)$ to define the code and thus we get a
$(9,6)$ convolutional code. The generator matrix is $\hat{f}(z) =
\di\sum_{i=0}^n \hat{\al_i}z^i$ where $\hat{\al_i}$ denotes the first
  $6$ rows of $\al_i$, considered as a matrix.  

The control matrix is obtained from $f(z)^{2}/z^{3t}$ using the last
$3$ columns of the $\al_i$.

\begin{lemma}\label{lem:hat} $\un{x} \hat{\al_i} + \un{y}\hat{\al_0}$
    has distance at least $1$ for $1\ti 6$ vectors $\un{x},\un{y}$
    with $\un{y} \not = \un{0}$.
\end{lemma}

\subsubsection{Specific examples for characteristic $3$}

Define $f(z) = \al + \al_0z+\al z^2$. Then $\hat{f}(z) = \hat{\al} + 
\hat{\al_0}z+\hat{\al} z^2$ is a convolutional $(9,6)$ code of free
distance $8$. 
\\ Define $f(z) = \al+ \al_0z + \al z^3 + \al z^4$. Then $\hat{f}(z) = \hat{\al}+
\hat{\al_0}z + \hat{\al} z^3 + \hat{\al} z^4$ defines a $(9,6)$
convolutional code which has free distance $11$.   

A result similar to \thmref{best} can also be proved.

Suppose now $\CC$ is a  cyclic $(n,k,d_1)$ code over the field $F$ of
characteristic $3$. Suppose also  that  the dual of
$\CC$, denoted $\hat{\CC}$, is an $(n,n-k,d_2)$ code. 

Let $d = min(d_1,d_2)$. Suppose 
$f(g)= \di\sum_{i=0}^r\be_ig^i$, with $\be_i \in F, (\be_r\not = 0)$, 
is a generating polynomial for $\C$. In $f(g)$, assume $\be_0 \not =
0$. 

Consider $f(z) = \di\sum_{i=1}^r \al_iz^i$ where now $\al_i = \be_i
\al$ with $\al$ as above in $F(C_3\ti C_3)$. Note that if $\be_i = 0$
then $\al_i = 0$. 
Replace some $\al_i$, say $\al_t$, by $\al_0$  (considered as
members of $F(C_3\ti C_3)$).

So assume  $f(z) = \di\sum_{i=0}^r \al_iz^i$ with this
$\al_t = \al_0$  and other  $\al_i = \be_i\al$ 
(for $i \not = t$). 

Then  $f(z)^3 = \be_t^3z^{3t}$ giving that $f(z)\ti (f(z)^2/(\be_t^3z^{3t})
= 1$. We now use 
$f(z)$ to generate a convolutional code by taking  the first 6 rows 
of the $\al_i$. Thus the generating matrix is $\hat{f}(z)=
\di\sum_{i=0}^r\hat{\al}_i\be_iz^i$ where  $\hat{\al}_i$ consists of the
first $6$ rows of  $\al$ for $i\not = t$ and $\hat{\al_t}$ consists of
the first $6$ rows of $\al_0$.

For the following theorem assume  the invertible element $\al_0$ does
not occur in the first or the last position of $f$.

\begin{theorem}\label{thm:best1} Let $\CC$ denote the convolutional code
  with generator
  matrix $\hat{f}$. Then the free distance of $\CC$ is at least $d+4$.
\end{theorem}

\section{General considerations} 
Suppose it is required that 
 a degree $n$ polynomial $f(z)= \al_0 +
\al_1 z + \al_2z^2 + \ldots +
\al_nz^n$ is to 
have an inverse in $R[z]$. Then sufficient conditions on the
$\al_i$ are obtained by formally multiplying $f(z)$
 by a general $g(z)$ and making
sure in the product that the
coefficient of $z^0$ is 1 and the coefficient of $z^i$ is 0 for $i >
0$.

If all the $\al_i$ commute (as for  group rings on abelian
groups),  $2\al_i =0$ (as in characteristic 2), and  $\al_0^2 = 1,
\al_i^2 = 0,
\forall i \geq 1, $ then $g(z) = \al_0 - \al_1 z -
\al_2z^2 - \ldots - \al_nz^n$  satisfies $f(z)g(z) =1$. Choosing 
different $\al_i$ will maximise the distance.

It is easy to obtain elements $\al$ in the group ring with $\al^2 =
0$.
Consider for example $\Z_2C_{2n}$. Then $w_i = g^i+g^{n+i}$ for $0\leq
i < n$ satisfy
$w_i^2 = 0$ and any combination $\al$ of the $w_i$ satisfies $\al^2 =
0$. It is then a matter of choosing  suitable 
combinations.
 
\subsection{Nilpotent type} Many group rings $R$ have elements 
$\al$ such that
$\al^n =0$ (and $\al^{r} \not = 0, r < n$). These can be exploited to
 produce convolutional codes.

\subsubsection{Example} Consider $\F C_{14}$ where $\F$ has
characteristic $2$.  Let $w_0 = 1+g^5+g^6 +g^{12}+
g^{13}, w_1 = 1+g^2+g^5+g^7+g^9+g^{12}, w_2 = 1+g+g^3+g^7+g^8+g^{10}$ 
and define $p = w_0+w_1z+w_2z^2$. Then $p^2 =
1$. Since $w_i^2=0$ for $i\geq 1$,   consider a rate of
 $\frac{1}{2}$. Thus consider the convolutional code with encoder
matrix obtained from the  first $7$ rows of $p$ and then the control
matrix is obtained from the last $7$ columns of $p$.

\subsection{Further examples} Consider $\Z_2C_8$ generated by $g$.
Define $u= \al_0 + (1+g^4)z+(1+g^2)z^2+(1+g)z^3$
where $\al_0^8=1$. There are a number of choices for $\al_0$,
e.g. $\al_0 = 1+g+g^3$.

Then $u^2 = \al_0^2 + (1+g^8)z + (1+g^4)z^2+(1+g^2)z^6= \al_0^2+
(1+g^4)z^2+(1+g^2)z^6 $, $u^4=\al_0^4+(1+g^4)z^{12}$ and $u^8 = 1$.

Then $u$ can be used to define a convolutional code. Now $1+g^4$ has
$\rank 4$ so for best results make it an $(8,4)$ convolutional code by
taking the first $4$ rows of the matrices of $u$. 

This is an $(8,4,9)$ convolutional code
with degree/memory $\delta = 6$.

The rate could be increased but this would reduce the contribution from
$(1+g^4)$ matrix to distance essentially $0$ as it has $\rank = 4$. 
This would give a $(8,6,7)$
convolutional code. 

To go further, consider  $\Z_2C_{16}$ etc.\ . Here use degree $6$ or $3$ as
the largest 
power of $z$ and it is then possible to get a  $(16, 8, 9)$ 
convolutional code. As  $\rank (1+g^4) = 8$ it is probably possible to
construct  a $(16,12,9)$ but details have not
been worked out.

These are binary codes. Going to bigger fields should give better
distances.

\section{Hamming type}\label{sec:hamm}

Set $R=\Z_2(C_4\ti C_2)$. Suppose $C_4$
is generated by $a$ and $C_2$ is generated by $h$. Consider
$\al_0 = 1+h(1+a^2)$ and $\al_i = 1+h(a+a^2+a^3)$ or $\al_i= 0$ for $i >0$. 
Then $\al_0^2 = 1$ and $\al_i^2 = 0$. Define $w(z)=
\di\sum_{i=0}^n \al_iz^i$ in  $RC_\infty$. By \thmref{ideal}, $w^2= 1$. 

Let $A = \left(\begin{array}{rrrr} 1&0&1&0\\0&1&0&1 \\1&0&1&0\\0&1&0&1
\end{array}\right)$, $B=  \left(\begin{array}{rrrr}
  0&1&1&1\\1&0&1&1  \\1&1&0&1\\1&1&1&0
\end{array}\right)$ and $I$ is the identity $4\ti 4$ matrix.
The matrix corresponding to $\al_0$ is then $\left(\begin{array}{rr}I
  & A \\ A&I\end{array}\right)$ and the matrix
  corresponding to  $\al_i$, $i\not = 0$, is either 
 $\left(\begin{array}{rr}I& B
  \\B&I\end{array}\right)$ or the zero matrix. 

Now specify that the first $ 4$ rows of $w$ formulate the generator
matrix of a code and then the last four columns of $w$ formulate the control
matrix. This gives a convolutional code of length $8$ and dimension
$4$. It is easy to transform the  resulting code into a  systematic code.

The generator matrix is $G(z) = (I,A) + \de_1(I,B)z + \de_2(I,B)z^2 + \ldots
+ \de_n(I,B)z^n$, where $\de_i \in \{0,1\}$. The control matrix is $H(z) =
\left(\begin{array}{c}A\\I\end{array}\right) +
  \de_1\left(\begin{array}{c}B\\I\end{array}\right)z
    +\de_2\left(\begin{array}{c}B\\I\end{array}\right)z^2  \ldots +  
\de_n\left(\begin{array}{c}B\\I\end{array}\right)z^n$.

The $(I,A)$ may be moved to the coefficient of any $z^i$ in which case
the (natural) control matrix will need to be divided by a power of $z$
to get the true control matrix. 

This convolutional code may be considered as a Hamming type
convolutional code as $(I,B)$ is a generator matrix of the Hamming
$(8,4)$ code.  

For $n=1$ the free distance turns out to be  $6$; this can be proved
in a similar manner to \thmref{freed}. 

\subsection{Example of this type}
$G(z) = (I,B) + (I,A)z + (I,B)z^2$ with  control matrix  $H(z)/z^2$
where $H(z) =
\left(\begin{array}{c}B\\I\end{array}\right) +
  \left(\begin{array}{c}A\\I\end{array}\right)z
    +\left(\begin{array}{c}B\\I\end{array}\right)z^2$ \, has free
      distance $10$.

\subsection{From cyclic to Hamming type}
For $n\geq 2$, proceed as previously to define the polynomials by
reference to corresponding cyclic linear polynomials. This will give
convolutional codes of this type of increasing free distance. Note
that $(I,A)$ has distance 
$2$, $(I,B)$ (the Hamming Code) has distance $4$, any combination
of $(I,A)$ and $(I,B)$ has distance $\geq 1$.

The following may be proved in a similar manner to \thmref{best}.

Suppose now $\CC$ is a  cyclic $(n,k,d_1)$ code over the field $F$ 
of characteristic $2$ and that  the dual of
$\CC$, $\hat{\CC}$, is an $(n,n-k,d_2)$ code. Let $d = min(d_1,d_2)$. 

Assume  $f(g) = \di\sum_{i=1}^r \be_ig^i$ is a generator polynomial
for $\CC$.
In $f(g)$, it is possible to arrange that $\be_0 \not = 0$ and
naturally assume that  
$\be_r \not = 0$. 
Define  $f(z) = \di\sum_{i=1}^r \al_iz^i$ with the $\al_i=
\be_i\al_i$, $i\not = t$ and $\al_t = \al_0$. 

Then  $f(z)^2 = z^{2t}$ giving $f(z)\ti f(z)/z^{2t} = 1$. Now use 
$f(z)$ to generate a convolutional code by taking just the first four 
rows of the $\al_i$. Thus the generating matrix is $G=
\di\sum_{i=0}^r\hat{\al}_iz^i$ where $\hat{\al}_i$ 
consists of  the first four rows of the matrix of $\al_i$.

\begin{theorem}\label{thm:hamming} $\CC$ has free distance at least $d+8$.
\end{theorem}

National University of Ireland, Galway

Galway

Ireland.
\end{document}